\documentclass[twocolumn,pra,showpacs,superscriptaddress,floatfix]{revtex4}
\usepackage{amsmath}
\usepackage{graphicx}
\usepackage{amssymb}
\usepackage{amsmath}

\newcommand{\Fn}[1]{\ensuremath{{\textrm{F}_{#1}}}}
\newcommand{\Gn}[1]{\ensuremath{{\textrm{G}_{#1}}}}
\newcommand{\Nn}[1]{\ensuremath{{\textrm{N}_{#1}}}}
\newcommand{\Wn}[1]{\ensuremath{{\textrm{W}_{#1}}}}
\newcommand{\NOT}{\textsc{not}}
\newcommand{\e}{{\textrm e}}
\newcommand{\I}{{\textrm i}}

\begin{document}
\title{Nested composite NOT gates for quantum computation}
\author{Jonathan A. Jones}\email{jonathan.jones@qubit.org}\thanks{Telephone +44 1865 272247, FAX: +44 1865 272400}
\affiliation{Centre for Quantum Computation, Clarendon Laboratory, University of Oxford, Parks Road, OX1 3PU, United Kingdom}
\date{\today}
\begin{abstract}
% Text of abstract
Composite pulses provide a simple means for constructing quantum logic gates which are robust to small errors in the control fields used to implement them.  Here I describe how antisymmetric composite \NOT\ gates can be nested to produce gates with arbitrary tolerance of errors.
\end{abstract}
\maketitle

% main text
%\section{Introduction}
Quantum information processing requires the ability to transform the state of physical systems representing quantum bits through a series of unitary transformations representing quantum logic gates \cite{Bennett2000}.  In real physical systems such gates will be vulnerable to systematic errors arising from imperfections in the control fields, and it is useful to design gates which are robust to such imperfections.  This can be achieved through composite pulses, sometimes called composite rotations \cite{Levitt1979,Levitt1986,Cummins2003,Jones2011,Merrill2012,Jones2013,Husain2013}.

Although many composite pulses have been developed in the context of nuclear magnetic resonance (NMR) experiments, their potential applications are much wider, including electron para\-mag\-netic resonance \cite{Morton2005a}, muon spin resonance \cite{Clayden2012}, ion traps \cite{Gulde2003}, neutral atoms \cite{Rakreungdet2009}, atom interferometers \cite{Butts2013}, and optical retarders \cite{Ardavan2007b,Ivanov2012}.  Note that for applications in quantum information it is necessary to use Class A composite pulses \cite{Levitt1986}, sometimes called general rotors, which correct imperfections in the underlying gate; simpler ``point-to-point'' pulses which only correct the effects of the gate on particular initial states are not suitable.

I will describe each rotation in a composite pulse by its propagator
\begin{equation}
\theta_\phi=\e^{-\I\theta\sigma_\phi/2}=\cos(\theta/2)\openone-\I\sin(\theta/2)\sigma_\phi
\end{equation}
\begin{equation}
\sigma_\phi=\cos\phi\,\sigma_x+\sin\phi\,\sigma_y
\end{equation}
where $\theta$ is the rotation angle and $\phi$, the pulse phase, fixes the rotation axis in the $xy$-plane.  As usual propagators must be written with time running from right to left, the reverse of the usual order for pulse sequences.  In the presence of pulse strength errors, when the amplitude of the driving field deviates from its nominal value by some fraction $\epsilon$, the rotation angle $\theta$ of each pulse is also increased by the same fraction.  In particular an error prone pulse can be written as
\begin{equation}
\pi'_\phi=[\pi(1+\epsilon)]_\phi=\pi_\phi(\epsilon\pi)_\phi
\end{equation}
so that the propagator for an error prone pulse can be written as a product of the desired and error components.

Here, as elsewhere \cite{Jones2013,Husain2013}, I will concentrate on attempts to construct \NOT\ gates, that is $\pi_0$ rotations, using only $\pi_\phi$ rotations, as composite pulses of this kind have many desirable properties.  Note that \NOT\ gates play an important role in experimental techniques such as dynamical decoupling \cite{Viola1999,Souza2011b,Souza2012}.  A number of relatively short composite pulses of this kind have been described, and it is known that some of these, most notably \Fn1, can be nested to produce longer pulse sequences with much higher error tolerance \cite{Husain2013}. Here I explain why nesting is successful for \Fn1 and for other antisymmetric composite pulses, but does not work for most other types of composite pulse.  I do not consider here the use of NMR phase cycles and supercycles \cite{Levitt1981,Shaka1987a}, which have proved useful in some NMR implementations of dynamical decoupling \cite{Souza2012}.

\section{The \Fn{1} composite $\pi$ pulse}
The \Fn1 composite pulse \cite{Wimperis1991} implements a $\pi_0$ rotation as a sequence of five $\pi$ pulses, with the propagator
\begin{align}
V_{\textrm{F}_1}&=\pi'_{\phi_5}\pi'_{\phi_4}\pi'_{\phi_3}\pi'_{\phi_2}\pi'_{\phi_1}\\
&=\pi_{\phi_5}\delta_{\phi_5}\pi_{\phi_4}\delta_{\phi_4}\pi_{\phi_3}\delta_{\phi_3}\pi_{\phi_2}\delta_{\phi_2}\pi_{\phi_1}\delta_{\phi_1}
\end{align}
where $\delta=\epsilon\pi$.  Using the identity
\begin{equation}
\theta_\beta\,\pi_\alpha=\pi_\alpha\,\theta_{2\alpha-\beta}
\label{eq:pimove}
\end{equation}
allows the $\pi$ pulses to be propagated through the sequence, in effect separating the evolution into two parts, the desired overall evolution and an error term
\begin{equation}
V_{\textrm{F}_1}=(\pi_{\phi_5}\pi_{\phi_4}\pi_{\phi_3}\pi_{\phi_2}\pi_{\phi_1})(\delta_{\phi'_5}\delta_{\phi'_4}\delta_{\phi'_3}\delta_{\phi'_2}\delta_{\phi'_1})
\end{equation}
where the desired evolution depends on the phases of the pulses as applied, but the error term depends on the phase of these pulses in the interaction frame \cite{Levitt1986} or toggling frame \cite{Suter1987a,Odedra2012b}, given by \cite{Jones2013}
\begin{equation}
\phi'_j=(-1)^{(j+1)}\phi_j+\sum_{k<j}(-1)^{(k+1)}2\phi_k. \label{eq:phitog}
\end{equation}

The \Fn1 composite pulse uses \textit{antisymmetric} pulse phases, so that $\phi_5=-\phi_1$, $\phi_4=-\phi_2$ and $\phi_3=0$, which has two important consequences.  Firstly the desired evolution term is always equal to a $\pi_0$ pulse, whatever values are chosen for $\phi_1$ and $\phi_2$, and these phases can be chosen to create error tolerance.  Secondly the toggling frame phases are \textit{symmetric}, and are given by
\begin{align}
\phi'_5=\phi'_1&=\phi_1\\
\phi'_4=\phi'_2&=2\phi_1-\phi_2\\
\phi'_3&=2\phi_1-2\phi_2.
\end{align}
This relationship between antisymmetric pulse phases and symmetric toggling frame phases underlies the success of antisymmetric composite \NOT\ gates.

To produce a good composite pulse requires simply that the error term be made as close as possible to the identity operation.  It is tempting to expand the error propagator as Taylor series in $\epsilon$ and then sequentially remove error terms, but it is more insightful to consider the underlying generator of the error propagator.  This can be written as
\begin{equation}
\delta_{\phi'_5}\delta_{\phi'_4}\delta_{\phi'_3}\delta_{\phi'_2}\delta_{\phi'_1}=\exp\{-\I(\Delta_1+\Delta_2+\dots)\}
\end{equation}
where the individual error terms can be calculated using the  Baker--Campbell--Hausdorff (BCH) relation \cite{Campbell1897,EBWbook} as
\begin{equation}
\Delta_1=\frac{\delta}{2}\sum_j\sigma_{\phi'_j}
\label{eq:del1}
\end{equation}
and
\begin{align}
\Delta_2&=-\frac{\I\delta^2}{8}\sum_j\sum_{k<j}\big[\sigma_{\phi'_j},\sigma_{\phi'_k}]\\
&=-\frac{\delta^2}{4}\sum_j\sum_{k<j}\sin(\phi'_j-\phi'_k)\,\sigma_z.\label{eq:del2}
\end{align}
Note that in NMR treatments it is customary to perform these calculations using average Hamiltonian theory based on the Magnus expansion  \cite{Haeberlen1968,Tycko1984,Tycko1985a,Odedra2012b,EBWbook}, and the numbering convention differs by one, so that these errors are referred to as zero and first order errors respectively.

From the form of Eq.~\ref{eq:del2} it can be seen that the second order error term will vanish for sequences such as \Fn1 which have symmetric toggling frame phases.  Thus if the phases can be chosen such that the first order error, Eq.~\ref{eq:del1}, is suppressed, then the resulting composite pulse will have third order error terms.  The corresponding propagator fidelity \cite{Jones2011}, which can in this case be calculated from the trace of the error propagator, will be correct up to sixth order infidelity.  This can be achieved by choosing $\phi_1=3\psi$ and $\phi_2=\psi$ with $\psi=\pm\arccos(-1/4)$, a result which can be understood geometrically by considering the first order error as a vector sum \cite{Jones2013}.  The \Fn1 pulse can be summarised by listing its phases
\begin{equation}
\mbox{\boldmath$(\phi)$}=(\phi_1,\, \phi_2,\, \phi_3,\, \phi_4,\, \phi_5)=(3\psi,\,\psi,\,0,\,-\psi,\,-3\psi)
\end{equation}
or the corresponding phases in the toggling frame
\begin{equation}
\mbox{\boldmath$(\phi')$}=(3\psi,\,5\psi,\,4\psi,\,5\psi,\,3\psi)
\end{equation}
where Eq.~\ref{eq:phitog} can be used to interconvert these two forms.  Note that the choice of plus or minus sign for $\psi$ is arbitrary, but must be made consistently.

To make further progress it is necessary to consider the form of the higher order error terms, $\Delta_3$ and so on.  Although the detailed forms of these higher terms swiftly become complicated, it is possible to make some general statements.  In particular, the vanishing of $\Delta_2$ is simply a specific example of a more general property, that all even order terms disappear in the expansion of a time symmetric sequence of operators \cite{Wang1972,Burum1981,Merrill2012}.  The third order term will be a sum of double commutators, and so will be a sum of $\sigma_\phi$ terms, and similarly for all odd-order terms, while the missing even order terms would be proportional to $\sigma_z$.  Thus all the error terms for the \Fn1 composite pulse lie in the $xy$-plane, a fact whose significance will soon become clear.

\section{The \Fn{n} family}
\Fn1 is simply the first member of a family \cite{Wimperis1991,Husain2013} of composite \NOT\ gates obtained by recursively nesting the \Fn1 pulse.  There has been considerable interest within the NMR community in designing high order robust pulses by recursive or iterative expansions \cite{Levitt1983,Tycko1984}, usually concentrating on the form of the error propagator \cite{Husain2013,Odedra2012b}. Naive iterative expansions are sometimes useful in the design of point-to-point pulses \cite{Tycko1984}, but less successful for Class A general rotors \cite{Husain2013}.

The \Fn1 composite pulse can, however, be extended by nesting with phase reversals.  (The first extension to obtain \Fn2 has been known for many years \cite{Wimperis1991}, and this was subsequently extended and generalised \cite{Husain2013}.)  The phase patterns are given by
\begin{equation}
\mbox{\boldmath$(\phi_{n+1})$}=(3\psi+\mbox{\boldmath$\phi_{n}$},\,\psi-\mbox{\boldmath$\phi_{n}$},\,\mbox{\boldmath$\phi_{n}$},\,-\psi-\mbox{\boldmath$\phi_{n}$},\,-3\psi+\mbox{\boldmath$\phi_{n}$})
\label{eq:Fnn}
\end{equation}
with $\mbox{\boldmath$(\phi_{0})$}=(0)$, so that \Fn1 is the sequence of 5 pulses described above, while \Fn2 is a sequence of 25 pulses.  The origin of this formula is more easily seen by considering the corresponding toggling frame phases
\begin{equation}
\mbox{\boldmath$(\phi'_{n+1})$}=(3\psi+\mbox{\boldmath$\phi'_{n}$},\,5\psi+\mbox{\boldmath$\phi'_{n}$},\,4\psi+\mbox{\boldmath$\phi'_{n}$},\,5\psi+\mbox{\boldmath$\phi'_{n}$},\,3\psi+\mbox{\boldmath$\phi'_{n}$})
\end{equation}
which makes clear that \Fn2 is obtained by applying the \Fn1 pattern to \Fn1 \textit{in the toggling frame}, and so on.

The error propagator for \Fn2 is most simply analysed by breaking it into five equal parts.  Each of these is equivalent to the error propagator for \Fn1, except that the overall phase of each error propagator is shifted by some angle.  The outer sequence of phases in the \Fn2 pulse cycles these phases such as to cause \textit{any} error terms in the $xy$-plane to be cancelled to first order, and the overall symmetry of the error propagator ensures that these terms will also be cancelled to second order.  The first error term which therefore survives in the \Fn2 error propagator is the ninth order term, arising from a third order combination of third order terms.  The next member of the family, \Fn3, can be analysed in precisely the same way, leaving only error terms of order 27 and higher.  At each stage all the remaining errors are odd-order and found in the $xy$-plane, and so are cancelled to third order by the next iteration.  Thus, as previously conjectured \cite{Husain2013}, the infidelity of a general \Fn{n} sequence is of order $2\times3^n$.

\section{More complex pulses}
As noted previously \cite{Husain2013} a similar pattern is seen for other antisymmetric composite $\pi$ pulses, such as the \Gn{n} family of targeted composite pulses, and the \Nn{n} family of narrowband pulses, as well as for combinations of these pulses in which different phase patterns are applied at each iteration.  The same behaviour is also seen for antisymmetric versions of the \Wn{n} family of $\pi$ pulses \cite{Husain2013}.  All these results are easily understood when the error propagator is examined in the toggling frame: each pattern of phases produces an error propagator which only has terms in the $xy$-plane, and each outer pattern modulates the size of inner error terms in the obvious way.  This view of nested pulses completely explains previous observations, such as the GF nested pulse having sixth order infidelity around $\epsilon=0$ and perfect fidelity at $\epsilon\approx\pm0.720$ \cite{Husain2013}.

%\section{Off-resonance errors}
A similar approach can be used to tackle off-resonance errors, which occur in experimental approaches such as NMR when the control field is not quite in resonance with the qubit transition.  In this case the error term for a single rotation is more complicated than for pulse strength errors \cite{Jones2013}, but the error term still lies in the $xy$-plane, with both the rotation angle and the phase of the error term being functions of the size of the off-resonance error, such that even-order error terms lie parallel to the main pulse, while odd-order terms lie perpendicular to it.

Correcting such errors requires a slightly more complex approach than for pulse strength errors\cite{Jones2013,Odedra2012b}, but can be achieved using longer composite pulses, such as the ASBO-9 family \cite{Odedra2012b}, which comprises nine $\pi$ pulses with antisymmetric phases and which corrects both pulse strength errors and off-resonance errors up to and including second order terms, so that the composite pulse has sixth order infidelity with respect to both errors.  As before, ASBO-9 pulses can be nested in the obvious way, resulting in a sequence of 81 $\pi$ pulses with infidelities of order 18 with respect to both errors.  As usual ASBO-9 can be combined with other composite pulses to get more complex patterns of error suppression.

\section{Symmetric pulses}
All the composite pulses considered so far are antisymmetric pulses, which leads to symmetric phases in the toggling frame.  Symmetric composite pulses frequently have better responses to simultaneous pulse strength and off-resonance errors \cite{Jones2013}, and it would be desirable to be able to nest such pulses to achieve higher error suppression.  This approach is not, however, successful.

%As an example c
Consider the symmetric five pulse sequence \cite{Jones2013}
\begin{equation}
\mbox{\boldmath$(\phi)$}=(\alpha,\,\beta,\,2\beta-2\alpha,\,\beta,\,\alpha)
\end{equation}
with pulse phases $\beta=2\alpha\pm\arccos[-(1+2\cos\alpha)/2]$ and $\alpha=\mp2\arcsin(\sqrt[4]{5/32})$, which suppresses both first and second order pulse strength errors.  The lowest surviving error term is a third order term in the $xy$-plane, and nesting the composite pulse will cancel this term.  However because the toggling frame phases are not symmetric (they are in fact antisymmetric in this case) the higher even order errors are not automatically suppressed, and so the nested sequence will contain a fourth order error term.  Similar arguments apply to other symmetric and asymmetric composite pulses.  This does not, of course, completely rule out the possibility of developing a recursive procedure for expanding such pulses, but does explain why the current approach will not lead there.

\section{Conclusions}
The previous conjecture \cite{Husain2013} that \Fn{n} composite pulses could be indefinitely nested using Eq.~\ref{eq:Fnn} has now been confirmed, as has the equivalent behaviour of other antisymmetric sequences of $\pi$ pulses.  These composite pulses have symmetric error phases in the toggling frame, leading to a cascading suppression of amplitude error terms.  The general \Fn{n} pulse contains $5^n$ component pulses and suppresses all error terms below order $3^n$.  Thus to achieve an error of the form $O(\epsilon^q)$ requires a pulse sequence of length about $q^p$ with $p=\ln5/\ln3\approx1.47$, much shorter than some previous estimates \cite{Brown2004}.  The same approach can be used to achieve simultaneous suppression of amplitude and off-resonance errors.  This result, however, is limited to composite $\pi$ pulses.

It has also been conjectured that the \Wn{n} approach \cite{Husain2013} allows composite pulses to be constructed with length linear in the desired error suppression, although only low members of the family were explicitly located.  Recently Low \textit{et al.} have shown \cite{Low2013} that higher members do indeed exist, although actually locating them remains a difficult problem.  Thus the nesting approach is still the only technique known for explicitly constructing arbitrary precision composite pulses.


\begin{thebibliography}{33}
\expandafter\ifx\csname natexlab\endcsname\relax\def\natexlab#1{#1}\fi
\expandafter\ifx\csname bibnamefont\endcsname\relax
  \def\bibnamefont#1{#1}\fi
\expandafter\ifx\csname bibfnamefont\endcsname\relax
  \def\bibfnamefont#1{#1}\fi
\expandafter\ifx\csname citenamefont\endcsname\relax
  \def\citenamefont#1{#1}\fi
\expandafter\ifx\csname url\endcsname\relax
  \def\url#1{\texttt{#1}}\fi
\expandafter\ifx\csname urlprefix\endcsname\relax\def\urlprefix{URL }\fi
\providecommand{\bibinfo}[2]{#2}
\providecommand{\eprint}[2][]{\url{#2}}

\bibitem[{\citenamefont{Bennett and DiVincenzo}(2000)}]{Bennett2000}
\bibinfo{author}{\bibfnamefont{C.~H.} \bibnamefont{Bennett}} \bibnamefont{and}
  \bibinfo{author}{\bibfnamefont{D.~P.} \bibnamefont{DiVincenzo}},
  \bibinfo{journal}{Nature} \textbf{\bibinfo{volume}{404}},
  \bibinfo{pages}{247} (\bibinfo{year}{2000}).

\bibitem[{\citenamefont{Levitt and Freeman}(1979)}]{Levitt1979}
\bibinfo{author}{\bibfnamefont{M.~H.} \bibnamefont{Levitt}} \bibnamefont{and}
  \bibinfo{author}{\bibfnamefont{R.}~\bibnamefont{Freeman}},
  \bibinfo{journal}{J. Magn. Reson.} \textbf{\bibinfo{volume}{33}},
  \bibinfo{pages}{473} (\bibinfo{year}{1979}).

\bibitem[{\citenamefont{Levitt}(1986)}]{Levitt1986}
\bibinfo{author}{\bibfnamefont{M.~H.} \bibnamefont{Levitt}},
  \bibinfo{journal}{Prog. NMR Spectrosc.} \textbf{\bibinfo{volume}{18}},
  \bibinfo{pages}{61} (\bibinfo{year}{1986}).

\bibitem[{\citenamefont{Cummins et~al.}(2003)\citenamefont{Cummins, Llewellyn,
  and Jones}}]{Cummins2003}
\bibinfo{author}{\bibfnamefont{H.~K.} \bibnamefont{Cummins}},
  \bibinfo{author}{\bibfnamefont{G.}~\bibnamefont{Llewellyn}},
  \bibnamefont{and} \bibinfo{author}{\bibfnamefont{J.~A.} \bibnamefont{Jones}},
  \bibinfo{journal}{Phys. Rev. A} \textbf{\bibinfo{volume}{67}},
  \bibinfo{pages}{042308} (\bibinfo{year}{2003}).

\bibitem[{\citenamefont{Jones}(2011)}]{Jones2011}
\bibinfo{author}{\bibfnamefont{J.~A.} \bibnamefont{Jones}},
  \bibinfo{journal}{Prog. NMR Spectrosc.} \textbf{\bibinfo{volume}{59}},
  \bibinfo{pages}{91} (\bibinfo{year}{2011}).

\bibitem[{\citenamefont{Merrill and Brown}(2012)}]{Merrill2012}
\bibinfo{author}{\bibfnamefont{J.~T.} \bibnamefont{Merrill}} \bibnamefont{and}
  \bibinfo{author}{\bibfnamefont{K.~R.} \bibnamefont{Brown}},
  \bibinfo{journal}{Advan. Chem. Phys.} \textbf{\bibinfo{volume}{(in press)}}
  (\bibinfo{year}{2012}), \urlprefix\url{http://arxiv.org/abs/1203.6392}.

\bibitem[{\citenamefont{Jones}(2013)}]{Jones2013}
\bibinfo{author}{\bibfnamefont{J.~A.} \bibnamefont{Jones}},
  \bibinfo{journal}{Phys. Rev. A} \textbf{\bibinfo{volume}{87}},
  \bibinfo{pages}{052317} (\bibinfo{year}{2013}).

\bibitem[{\citenamefont{Husain et~al.}(2013)\citenamefont{Husain, Kawamura, and
  Jones}}]{Husain2013}
\bibinfo{author}{\bibfnamefont{S.}~\bibnamefont{Husain}},
  \bibinfo{author}{\bibfnamefont{M.}~\bibnamefont{Kawamura}}, \bibnamefont{and}
  \bibinfo{author}{\bibfnamefont{J.~A.} \bibnamefont{Jones}},
  \bibinfo{journal}{J. Magn. Reson.} \textbf{\bibinfo{volume}{230}},
  \bibinfo{pages}{145} (\bibinfo{year}{2013}).

\bibitem[{\citenamefont{Morton et~al.}(2005)\citenamefont{Morton, Tyryshkin,
  Ardavan, Porfyrakis, Lyon, and Briggs}}]{Morton2005a}
\bibinfo{author}{\bibfnamefont{J.~J.~L.} \bibnamefont{Morton}},
  \bibinfo{author}{\bibfnamefont{A.~M.} \bibnamefont{Tyryshkin}},
  \bibinfo{author}{\bibfnamefont{A.}~\bibnamefont{Ardavan}},
  \bibinfo{author}{\bibfnamefont{K.}~\bibnamefont{Porfyrakis}},
  \bibinfo{author}{\bibfnamefont{S.~A.} \bibnamefont{Lyon}}, \bibnamefont{and}
  \bibinfo{author}{\bibfnamefont{G.~A.~D.} \bibnamefont{Briggs}},
  \bibinfo{journal}{Phys. Rev. Lett} \textbf{\bibinfo{volume}{95}},
  \bibinfo{pages}{200501} (\bibinfo{year}{2005}).

\bibitem[{\citenamefont{Clayden et~al.}(2012)\citenamefont{Clayden, Cottrell,
  and McKenzie}}]{Clayden2012}
\bibinfo{author}{\bibfnamefont{N.~J.} \bibnamefont{Clayden}},
  \bibinfo{author}{\bibfnamefont{S.~P.} \bibnamefont{Cottrell}},
  \bibnamefont{and} \bibinfo{author}{\bibfnamefont{I.}~\bibnamefont{McKenzie}},
  \bibinfo{journal}{J. Magn. Reson.} \textbf{\bibinfo{volume}{214}},
  \bibinfo{pages}{144} (\bibinfo{year}{2012}).

\bibitem[{\citenamefont{Gulde et~al.}(2003)\citenamefont{Gulde, Riebe,
  Lancaster, Becher, Eschner, Haffner, Schmidt-Kaler, Chuang, and
  Blatt}}]{Gulde2003}
\bibinfo{author}{\bibfnamefont{S.}~\bibnamefont{Gulde}},
  \bibinfo{author}{\bibfnamefont{M.}~\bibnamefont{Riebe}},
  \bibinfo{author}{\bibfnamefont{G.~P.~T.} \bibnamefont{Lancaster}},
  \bibinfo{author}{\bibfnamefont{C.}~\bibnamefont{Becher}},
  \bibinfo{author}{\bibfnamefont{J.}~\bibnamefont{Eschner}},
  \bibinfo{author}{\bibfnamefont{H.}~\bibnamefont{Haffner}},
  \bibinfo{author}{\bibfnamefont{F.}~\bibnamefont{Schmidt-Kaler}},
  \bibinfo{author}{\bibfnamefont{I.~L.} \bibnamefont{Chuang}},
  \bibnamefont{and} \bibinfo{author}{\bibfnamefont{R.}~\bibnamefont{Blatt}},
  \bibinfo{journal}{Nature} \textbf{\bibinfo{volume}{421}}, \bibinfo{pages}{48}
  (\bibinfo{year}{2003}).

\bibitem[{\citenamefont{Rakreungdet et~al.}(2009)\citenamefont{Rakreungdet,
  Lee, Lee, Mischuck, Montano, and Jessen}}]{Rakreungdet2009}
\bibinfo{author}{\bibfnamefont{W.}~\bibnamefont{Rakreungdet}},
  \bibinfo{author}{\bibfnamefont{J.~H.} \bibnamefont{Lee}},
  \bibinfo{author}{\bibfnamefont{K.~F.} \bibnamefont{Lee}},
  \bibinfo{author}{\bibfnamefont{B.~E.} \bibnamefont{Mischuck}},
  \bibinfo{author}{\bibfnamefont{E.}~\bibnamefont{Montano}}, \bibnamefont{and}
  \bibinfo{author}{\bibfnamefont{P.~S.} \bibnamefont{Jessen}},
  \bibinfo{journal}{Phys. Rev. A} \textbf{\bibinfo{volume}{79}},
  \bibinfo{pages}{022316} (\bibinfo{year}{2009}).

\bibitem[{\citenamefont{Butts et~al.}(2013)\citenamefont{Butts, Kotru, Joseph
  M.~Kinast, Radojevic, Timmons, and Stoner}}]{Butts2013}
\bibinfo{author}{\bibfnamefont{D.~L.} \bibnamefont{Butts}},
  \bibinfo{author}{\bibfnamefont{K.}~\bibnamefont{Kotru}},
  \bibinfo{author}{\bibfnamefont{J.~M.} \bibnamefont{Joseph M.~Kinast}},
  \bibinfo{author}{\bibfnamefont{A.~M.} \bibnamefont{Radojevic}},
  \bibinfo{author}{\bibfnamefont{B.~P.} \bibnamefont{Timmons}},
  \bibnamefont{and} \bibinfo{author}{\bibfnamefont{R.~E.}
  \bibnamefont{Stoner}}, \bibinfo{journal}{J. Opt. Soc. Am. B}
  \textbf{\bibinfo{volume}{30}}, \bibinfo{pages}{922} (\bibinfo{year}{2013}).

\bibitem[{\citenamefont{Ardavan}(2007)}]{Ardavan2007b}
\bibinfo{author}{\bibfnamefont{A.}~\bibnamefont{Ardavan}},
  \bibinfo{journal}{New J. Phys.} \textbf{\bibinfo{volume}{9}},
  \bibinfo{pages}{24} (\bibinfo{year}{2007}).

\bibitem[{\citenamefont{Ivanov et~al.}(2012)\citenamefont{Ivanov, Rangelov,
  Vitanov, Peters, and Halfmann}}]{Ivanov2012}
\bibinfo{author}{\bibfnamefont{S.~S.} \bibnamefont{Ivanov}},
  \bibinfo{author}{\bibfnamefont{A.~A.} \bibnamefont{Rangelov}},
  \bibinfo{author}{\bibfnamefont{N.~V.} \bibnamefont{Vitanov}},
  \bibinfo{author}{\bibfnamefont{T.}~\bibnamefont{Peters}}, \bibnamefont{and}
  \bibinfo{author}{\bibfnamefont{T.}~\bibnamefont{Halfmann}},
  \bibinfo{journal}{J. Opt. Soc. Am. A} \textbf{\bibinfo{volume}{29}},
  \bibinfo{pages}{265} (\bibinfo{year}{2012}).

\bibitem[{\citenamefont{Viola et~al.}(1999)\citenamefont{Viola, Knill, and
  Lloyd}}]{Viola1999}
\bibinfo{author}{\bibfnamefont{L.}~\bibnamefont{Viola}},
  \bibinfo{author}{\bibfnamefont{E.}~\bibnamefont{Knill}}, \bibnamefont{and}
  \bibinfo{author}{\bibfnamefont{S.}~\bibnamefont{Lloyd}},
  \bibinfo{journal}{Phys. Rev. Lett.} \textbf{\bibinfo{volume}{82}},
  \bibinfo{pages}{2417} (\bibinfo{year}{1999}).

\bibitem[{\citenamefont{Souza et~al.}(2011)\citenamefont{Souza, \'Alvarez, and
  Suter}}]{Souza2011b}
\bibinfo{author}{\bibfnamefont{A.~M.} \bibnamefont{Souza}},
  \bibinfo{author}{\bibfnamefont{G.~A.} \bibnamefont{\'Alvarez}},
  \bibnamefont{and} \bibinfo{author}{\bibfnamefont{D.}~\bibnamefont{Suter}},
  \bibinfo{journal}{Phys. Rev. Lett.} \textbf{\bibinfo{volume}{106}},
  \bibinfo{pages}{240501} (\bibinfo{year}{2011}).

\bibitem[{\citenamefont{Souza et~al.}(2012)\citenamefont{Souza, \'Alvarez, and
  Suter}}]{Souza2012}
\bibinfo{author}{\bibfnamefont{A.~M.} \bibnamefont{Souza}},
  \bibinfo{author}{\bibfnamefont{G.~A.} \bibnamefont{\'Alvarez}},
  \bibnamefont{and} \bibinfo{author}{\bibfnamefont{D.}~\bibnamefont{Suter}},
  \bibinfo{journal}{Phil. Trans. Roy. Soc. A} \textbf{\bibinfo{volume}{370}},
  \bibinfo{pages}{4748} (\bibinfo{year}{2012}).

\bibitem[{\citenamefont{Levitt and Freeman}(1981)}]{Levitt1981}
\bibinfo{author}{\bibfnamefont{M.~H.} \bibnamefont{Levitt}} \bibnamefont{and}
  \bibinfo{author}{\bibfnamefont{R.}~\bibnamefont{Freeman}},
  \bibinfo{journal}{J. Magn. Reson.} \textbf{\bibinfo{volume}{43}},
  \bibinfo{pages}{502 } (\bibinfo{year}{1981}).

\bibitem[{\citenamefont{Shaka and Keeler}(1987)}]{Shaka1987a}
\bibinfo{author}{\bibfnamefont{A.}~\bibnamefont{Shaka}} \bibnamefont{and}
  \bibinfo{author}{\bibfnamefont{J.}~\bibnamefont{Keeler}},
  \bibinfo{journal}{Prog. NMR Spectrosc.} \textbf{\bibinfo{volume}{19}},
  \bibinfo{pages}{47 } (\bibinfo{year}{1987}).

\bibitem[{\citenamefont{Wimperis}(1991)}]{Wimperis1991}
\bibinfo{author}{\bibfnamefont{S.}~\bibnamefont{Wimperis}},
  \bibinfo{journal}{J. Magn. Reson} \textbf{\bibinfo{volume}{93}},
  \bibinfo{pages}{199} (\bibinfo{year}{1991}).

\bibitem[{\citenamefont{Suter and Pines}(1987)}]{Suter1987a}
\bibinfo{author}{\bibfnamefont{D.}~\bibnamefont{Suter}} \bibnamefont{and}
  \bibinfo{author}{\bibfnamefont{A.}~\bibnamefont{Pines}}, \bibinfo{journal}{J.
  Magn. Reson.} \textbf{\bibinfo{volume}{75}}, \bibinfo{pages}{509 }
  (\bibinfo{year}{1987}).

\bibitem[{\citenamefont{Odedra et~al.}(2012)\citenamefont{Odedra, Thrippleton,
  and Wimperis}}]{Odedra2012b}
\bibinfo{author}{\bibfnamefont{S.}~\bibnamefont{Odedra}},
  \bibinfo{author}{\bibfnamefont{M.~J.} \bibnamefont{Thrippleton}},
  \bibnamefont{and} \bibinfo{author}{\bibfnamefont{S.}~\bibnamefont{Wimperis}},
  \bibinfo{journal}{J. Magn. Reson.} \textbf{\bibinfo{volume}{225}},
  \bibinfo{pages}{81 } (\bibinfo{year}{2012}).

\bibitem[{\citenamefont{Campbell}(1897)}]{Campbell1897}
\bibinfo{author}{\bibfnamefont{J.~E.} \bibnamefont{Campbell}},
  \bibinfo{journal}{Proc. Lond. Math. Soc.} \textbf{\bibinfo{volume}{29}},
  \bibinfo{pages}{14} (\bibinfo{year}{1897}).

\bibitem[{\citenamefont{Ernst et~al.}(1987)\citenamefont{Ernst, Bodenhausen,
  and Wokaun}}]{EBWbook}
\bibinfo{author}{\bibfnamefont{R.~R.} \bibnamefont{Ernst}},
  \bibinfo{author}{\bibfnamefont{G.}~\bibnamefont{Bodenhausen}},
  \bibnamefont{and} \bibinfo{author}{\bibfnamefont{A.}~\bibnamefont{Wokaun}},
  \emph{\bibinfo{title}{Principles of Nuclear Magnetic Resonance in One and Two
  Dimensions}} (\bibinfo{publisher}{Oxford University Press},
  \bibinfo{year}{1987}).

\bibitem[{\citenamefont{Haeberlen and Waugh}(1968)}]{Haeberlen1968}
\bibinfo{author}{\bibfnamefont{U.}~\bibnamefont{Haeberlen}} \bibnamefont{and}
  \bibinfo{author}{\bibfnamefont{J.~S.} \bibnamefont{Waugh}},
  \bibinfo{journal}{Phys. Rev.} \textbf{\bibinfo{volume}{175}},
  \bibinfo{pages}{453} (\bibinfo{year}{1968}).

\bibitem[{\citenamefont{Tycko and Pines}(1984)}]{Tycko1984}
\bibinfo{author}{\bibfnamefont{R.}~\bibnamefont{Tycko}} \bibnamefont{and}
  \bibinfo{author}{\bibfnamefont{A.}~\bibnamefont{Pines}},
  \bibinfo{journal}{Chem. Phys. Lett.} \textbf{\bibinfo{volume}{111}},
  \bibinfo{pages}{462 } (\bibinfo{year}{1984}).

\bibitem[{\citenamefont{Tycko et~al.}(1985)\citenamefont{Tycko, Pines, and
  Guckenheimer}}]{Tycko1985a}
\bibinfo{author}{\bibfnamefont{R.}~\bibnamefont{Tycko}},
  \bibinfo{author}{\bibfnamefont{A.}~\bibnamefont{Pines}}, \bibnamefont{and}
  \bibinfo{author}{\bibfnamefont{J.}~\bibnamefont{Guckenheimer}},
  \bibinfo{journal}{J. Chem. Phys.} \textbf{\bibinfo{volume}{83}},
  \bibinfo{pages}{2775} (\bibinfo{year}{1985}).

\bibitem[{\citenamefont{Wang and Ramshaw}(1972)}]{Wang1972}
\bibinfo{author}{\bibfnamefont{C.~H.} \bibnamefont{Wang}} \bibnamefont{and}
  \bibinfo{author}{\bibfnamefont{J.~D.} \bibnamefont{Ramshaw}},
  \bibinfo{journal}{Phys. Rev. B} \textbf{\bibinfo{volume}{6}},
  \bibinfo{pages}{3253} (\bibinfo{year}{1972}).

\bibitem[{\citenamefont{Burum}(1981)}]{Burum1981}
\bibinfo{author}{\bibfnamefont{D.~P.} \bibnamefont{Burum}},
  \bibinfo{journal}{Phys. Rev. B} \textbf{\bibinfo{volume}{24}},
  \bibinfo{pages}{3684} (\bibinfo{year}{1981}).

\bibitem[{\citenamefont{Levitt and Ernst}(1983)}]{Levitt1983}
\bibinfo{author}{\bibfnamefont{M.~H.} \bibnamefont{Levitt}} \bibnamefont{and}
  \bibinfo{author}{\bibfnamefont{R.}~\bibnamefont{Ernst}}, \bibinfo{journal}{J.
  Magn. Reson.} \textbf{\bibinfo{volume}{55}}, \bibinfo{pages}{247 }
  (\bibinfo{year}{1983}).

\bibitem[{\citenamefont{Brown et~al.}(2004)\citenamefont{Brown, Harrow, and
  Chuang}}]{Brown2004}
\bibinfo{author}{\bibfnamefont{K.~R.} \bibnamefont{Brown}},
  \bibinfo{author}{\bibfnamefont{A.~W.} \bibnamefont{Harrow}},
  \bibnamefont{and} \bibinfo{author}{\bibfnamefont{I.~L.}
  \bibnamefont{Chuang}}, \bibinfo{journal}{Phys. Rev. A}
  \textbf{\bibinfo{volume}{70}}, \bibinfo{pages}{052318}
  (\bibinfo{year}{2004}).

\bibitem[{\citenamefont{Low et~al.}(2013)\citenamefont{Low, Yoder, and
  Chuang}}]{Low2013}
\bibinfo{author}{\bibfnamefont{G.~H.} \bibnamefont{Low}},
  \bibinfo{author}{\bibfnamefont{T.~J.} \bibnamefont{Yoder}}, \bibnamefont{and}
  \bibinfo{author}{\bibfnamefont{I.~L.} \bibnamefont{Chuang}}
  (\bibinfo{year}{2013}), \urlprefix\url{http://arxiv.org/abs/1307.2211}.

\end{thebibliography}
\end{document}